# New Physics with MeV Neutrino Sources Brighter than A Thousand Suns

Sanjib Kumar Agarwalla[1,2,#] and R. S. Raghavan[2,*]

[1]IFIC, University of Valencia & CSIC, Valencia, Spain, [2]Physics Dept, Virginia Tech, Blacksburg, USA

Short baseline oscillations of neutrinos ($\nu_e$ and $\tilde{\nu}_e$) due to active-sterile (a-s) mixing can be observed explicitly using MeV $\nu$ beams and existing/planned $\nu$ detectors. The typical baseline/energy (L/E) of this approach allows flavor survival waves to be observed in the spatial distribution of events inside the detector itself. Single/multiple oscillations can be tested using a variety of sources of $\nu_e$ and $\tilde{\nu}_e$ matched to Cerenkov (C), liquid scintillator (LS) and LENS (In-LS) detectors. Distinct tags for $\nu_e$ (from In) and $\tilde{\nu}_e$ (from p in the LS) in LENS allow access to $\nu_e$ and $\tilde{\nu}_e$ for probing (a-s) mixing, CPT symmetry and in a new way, limits on lepton number violation (LNV) via wrong helicity $\nu$ reactions comparable to limits for $0\nu\beta\beta$ decay via right handed currents.

Recent results on oscillations of high energy neutrinos ($\nu_e$) and antineutrinos ($\tilde{\nu}_e$) in MiniBooNE, MINOS and the LSND experiments present a somewhat confusing but actively discussed picture[1] of active-sterile (a-s) mixing and hints of CPT violation[2,3]. A recent analysis concludes that a resolution is unlikely with simple increase of statistics in these experiments[4]. In this Letter we discuss new strategies for progress in the problem based on explicit proof of oscillations in short baselines.

The basic factor in $\nu$ oscillations is L/E, the ratio of the baseline to the $\nu$ energy. Lower energies of a few MeV compared to GeV enable shorter baselines thus also, oscillations directly observable in spatial event distributions *within* a suitably matched detector. This idea was first pointed out for the Low Energy Neutrino Spectroscopy (LENS) detector[5] (L~2-3m, $\Delta$L~ 2.5cm) for $\nu_e$ oscillations at $E_\nu$ = 0.75 MeV (an idea picked up later for high energy experiments[4,6]). We now generalize this method by extending it to both $\nu_e$ and $\tilde{\nu}_e$ in LENS and in large scale detectors such as SuperKamiokande (SK) (L~60m, $\Delta$L ~1 m) and LENA. Indeed, the new physics accessible, particularly to LENS, is well beyond just (a-s) mixing.

Wide experimental choices are available for this program with MeV $\tilde{\nu}_e$ and $\nu_e$ sources and present or planned detectors. The development and cost of the approach is arguably much less than that for high energy experiments[4,7]. We illustrate direct observation of $\nu_e$ and $\tilde{\nu}_e$ oscillations in this approach in a model scenario and evaluate general sensitivities of experimental options over wide parameter ranges.

MeV $\nu_e$ from $\beta$-decay of $^8$B in the sun, studied for decades, have revolutionized neutrino physics. Such beams of high intensity ("brighter than a thousand suns") can be produced by "off the shelf" nuclear physics devices. Neutrinos from $\beta^\pm$ and electron capture decay after a large Lorentz boost ($\gamma$ >30) (beta-beams) have been proposed before.[8] In our approach $\gamma$ = 1 and sources are in situ in the target itself, eliminating the expensive and technically difficult extraction and boost parts and maximizing $\nu$ fluence. We consider SK(Gd[9] Cerenkov), LENA[10] (LS), and LENS (In-LS) detectors with coincidence *tags* for CC $\nu_e$ and $\tilde{\nu}_e$ capture.

Several built-in features in the strategy are attractive: 1) the intense $\nu$ beams with the technology of internal sources (see below) stress high statistics; 2) only established probes and interactions in the nuclear regime are involved, not GeV probes, often with less than fully understood low energy aspects; 3) the MeV $\nu$ spectra are defined precisely; 4) detection cross-sections are known; 5) the $\nu_e/\tilde{\nu}_e$ beams are intrinsically flavor pure; 6) $\nu_e/\tilde{\nu}_e$ beams and their specific tags can probe LNV via wrong-helicity cross reactions.

The detectors studied here are dimensionally matched to $P_{ee}$ waves with a wide range of frequencies. Unlike the search for a disappearance deficit using multiple detectors with multiple baselines we seek the $P_{ee}$ wave itself in a single detector. We illustrate this effect for specific model scenarios of oscillations[11].

Table 1 lists a selection of $\nu_e$ and $\tilde{\nu}_e$ sources that $\beta^\pm$ -decay with ~MeV energies produced by well known low energy machines of nuclear science. The on- line $\nu$ rates in Table 1 are ~$10^{22}$ to $10^{24}$/y. $^{51}$Cr can be made off-line via n-activation in a high-flux reactor[5,12] as well as on-line via a neutron generator based on the d+t$\rightarrow\alpha$+n (D-T) reaction. The D-T cross section suggests that ~$10^{15}$ n/s from 1A at 100 kV may be possible. Production of not only $^{51}$Cr ($\nu_e$) (via $^{52}$Cr (n,2n)) but, in principle, simultaneously also $^6$He ($\tilde{\nu}_e$) (via $^9$Be(n,$\alpha$)) is possible by the same initial n-fluence in a target assembly with concentric envelopes of Cr and Be. The n-fluence could be multiplied ($\rightarrow 10^{16}$ n/s) using internal subcritical fission in the n-generator[13] allowing on-line $^{51}$Cr to approach off-line strengths. Pending such progress, we use fluences from offline $^{51}$Cr. Neutrons from $^{12}$C(p,n)$^{12}$N can replace a n-generator. The reactions for $\nu_e/\tilde{\nu}_e$ with different

Table 1 Options for $\nu_e$ and $\tilde{\nu}_e$ beams for new physics

| Source | Mean Life (s) | Prod. Reaction (MeV) | Power (kW) | Target Mass | σ (b) | E(ν) (MeV) | ν /y @ Source |
|---|---|---|---|---|---|---|---|
| $^8$Li ($\tilde{\nu}_e$) | 1.21 | $^7$Li (d,p) (3) | (300) | 50mg | 0.4 | 0-13.1 | $3 \times 10^{22}$ |
| $^6$He ($\tilde{\nu}_e$) | 1.17 | $^9$Be(n,α) (14) | d+t → α+n (50) (also $^{12}$C(p,n)) | 300 kg | 0.1 | 0-3.5 | $2 \times 10^{22}$ |
| $^{12}$N ($\nu_e$) | 0.016 | $^{12}$C(p,n) (25) | (375) | 0.7g | 0.1 | 0-16.3 | $10^{22}$ |
| $^{51}$Cr ($\nu_e$) | $3.45 \times 10^6$ | $^{50}$Cr(n,γ) ($n_{th}$) $^{52}$Cr(n,2n) (14) | Nuc. Reactor d+t → α+n (50) | 50 kg 500 kg | 16 0.6 | 0.753 | $10^{24}$/10MCi $10^{23}$/y |

Table 2. Tagged CC reactions of $\tilde{\nu}_e$ and $\nu_e$ and detection technologies

| Tgt | Reaction | Σ (cm$^2$) | Tech | γ Tag (MeV) | Delay τ (μs) | ΔL (cm) @10MeV | $L_{max}$ (m) | E (Eff) | $E_{thresh}$ (MeV) |
|---|---|---|---|---|---|---|---|---|---|
| p | $\tilde{\nu}_e$ p → $e^+$ n (delay) γ | $6 \times 10^{-42}$ @10 MeV | C (SK) LS | 8 (Gd) 2.2 (p) | ~20 ~200 | 100 5 | ~60 | 0.67 0.76 | 5 2 |
| $^{115}$In | $\nu_e$ $^{115}$In → $e^-$ $^{115}$Sn* (delay)γ | $3 \times 10^{-44}$ @0.75 MeV | LENS (In) | 0.613 | 4.7 | 2.5 (@1 MeV) | ~3 | 1 | 0.11 |

specifics in Table 1 offer optimal choices for specific objectives (a-s mixing, LNV, CPT symmetry).

The possible $\nu_e$/$\tilde{\nu}_e$ reactions and figures of merit of the detectors are listed in Table 2. We consider only CC $\nu_e$/$\tilde{\nu}_e$ reactions that offer a coincidence tag to specify the energetics and discriminate strongly against background. The dominant $\tilde{\nu}_e$ reaction $\tilde{\nu}_e$+p → $e^+$+n→delay→+Gd→γ (8 MeV) may be possible in SK with Gd in the water[9]. In LS detectors such as LENS(In-LS) and LENA this reaction can be tagged using (np) capture yielding 2.2 MeV γ's. In addition to $\tilde{\nu}_e$ reactions, we stress the case for $\nu_e$ from the CPT point of view since only $\tilde{\nu}_e$ results from the Bugey experiment[14] are known so far.

The CC $\nu_e$ capture in $^{115}$In with LENS can be tagged via: $\nu_e$ + $^{115}$In → $e^-$ + $^{115}$Sn*+(delay τ) +γ with a threshold of 0.114 MeV.[15] It is the basis for detecting pp and other low energy solar neutrinos[16]. The signal events are picked out by the coincidence with the 0.613 γ cascade of $^{115}$Sn following $\nu_e$ capture after isomeric delay of ~4.76 μs. The high granularity of LENS (~$10^5$ cells for low energy events) is ideal for observing $P_{ee}$ waves with MCi off-line $^{51}$Cr as first proposed in ref. 5.

We now generalize this option using *on-line* source technology particularly with internal geometries and to other detectors. The projectile beam from an external accelerator can be transported to the reaction target at the center of the detector assembly (similar to that already present in SK for calibration). The 4π geometry results in high event rates in spite of the relatively small detector mass such as LENS. The powerful $^{115}$In or ($\tilde{\nu}_e$,p) tag is the key to suppressing background inherent in such strong internal sources[5]. They are shielded (by >$10^{14}$) against γ's and neutrons by~10 cm of heavy metal. The In tag is itself designed to suppress (by x$10^{12}$) the random internal background due to the radioactive $^{115}$In target.

The LENS detector is Indium loaded LS (a range of 8-30 wt % loading is possible[17]). The 3-D $P_{ee}$ wave can be observed in a (conceptually) spherical, modular assembly enveloping the source. $^{51}$Cr (EC decay) emits a mono-energetic $\nu_e$ while $^6$He emits a $\beta^-$ spectrum (0-3.5 MeV $\tilde{\nu}_e$) and $^{12}$N a $\beta^+$ specttum (0-16.3 MeV $\tilde{\nu}_e$). They can be measured concurrently by tags on $\nu_e$ from $^{51}$Cr and $^{12}$N (using the $^{115}$In tag) and on $\tilde{\nu}_e$ from $^6$He (using the ($\tilde{\nu}_e$,p) tag of protons of the LS).

Our focus is observing *explicit oscillations*. Standard simulations of events in the detector tagged by their spatial location are fitted[18] (with a routine 5% systematic error) to $P_{ee}$ waves in the chosen oscillation scenario. Fig 1 (top) illustrates $P_{ee}$ wave patterns (no oscillation/oscillation ratios) for $\nu_e$ and $\tilde{\nu}_e$ in LENS using parameters of ν mass models from best fit global data[11]. The effective L depends on the In loading. The presence of 2 frequencies in the 3+2 model is evident. With a built-in resolution ΔL ~2.5 cm[17], LENS can observe fast $\nu_e$ oscillations with $\Delta m^2$ as large as ~50-100 eV$^2$ with a $^{51}$Cr source at $E_\nu$ ~0.75 MeV.

The ν fluences and signal rates in available detectors (Table 3) can provide evidence of (a-s) mixing and CPT symmetry in ~1 year with 99% c.l. Fig. 1 (top) shows data patterns for LENS (30 wt.% In) for $\nu_e$ and and $\tilde{\nu}_e$ and bottom panel is for $\tilde{\nu}_e$ in SK(Gd) and a generic ~22.5 kton LS detector (e.g., LENA) with similar dimensions. The LENS geometry assumes an internal source while for SK and LS an external source at 20m from the top is assumed.

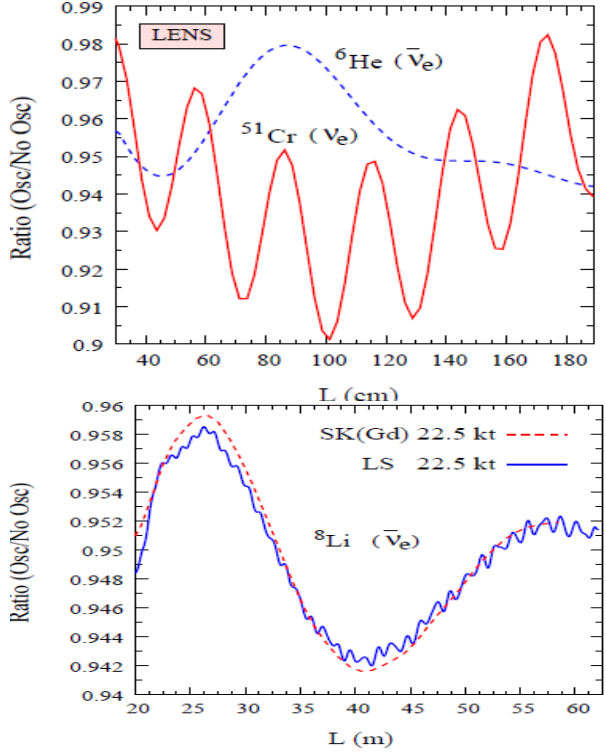

Fig. 1 $P_{ee}$ (a-s) oscillations (3+2 model[11]). Top panel: $\nu_e$ ($^{51}$Cr) in LENS-In (30 wt.% In) and $\tilde{\nu}_e$ ($^6$He) in LENS-LS (p). Bottom panel: $\tilde{\nu}_e$ ($^8$Li) in SK(Gd), LENA (LS).

Table 3 Unoscillated & oscillated events (3+2 model[11]) using 30 wt.% In (10 t) in LENS (internal source) and 22.5 kton SK(Gd) and LS (external source at 20m). The data can exclude no oscillation at 99% c.l.

| Detector | Source | ν Fluence | Unosc | Osc |
|---|---|---|---|---|
| LENS | $^{51}$Cr $\nu_e$ | $1.1 \times 10^{24}$ | 10190 | 9590 |
| LENS | $^6$He $\tilde{\nu}_e$ | $4.1 \times 10^{21}$ | 45070 | 43100 |
| SK(Gd) | $^8$Li $\tilde{\nu}_e$ | $7.0 \times 10^{21}$ | 128770 | 122570 |
| LS | $^8$Li $\tilde{\nu}_e$ | $5.8 \times 10^{21}$ | 161160 | 153320 |

Fig. 2 shows sensitivity limits on $P_{ee}$ waves in all three detectors compared to $\tilde{\nu}_e$ results from Bugey-Chooz (these limits will be improved in forthcoming reactor experiments such as Double Chooz, Daya Bay, RENO). Fig. 2 indicates high sensitivity to (a-s) mixing of $\tilde{\nu}_e$ from $^6$He in LENS and a complementary $^8$Li in

SK/LS. The $^6$He (LENS) data are thus competitive with proposed high energy schemes.

The SK/LS data here assume *external* $^8$Li sources. If these detectors are adapted to the *internal* source technique, sensitivity can approach $\sin^2 2\theta < 10^{-3}$ for $\tilde{\nu}_e$. The $\nu_e$ data from $^{51}$Cr is tentative, pending final design of the on-line n-source. The present work uses (as in Fig. 1) the fluence from a reactor source. While the sensitivity is less, it promises arguably, the first data for $\nu_e$ thus, it is of key relevance to CPT questions

The $\nu_e$ and $\tilde{\nu}_e$ beams in LENS with exclusive tags opens the opportunity to set limits on cross reactions e.g., capture of $\nu_e$ observed by the ($\tilde{\nu}_e$,p) tag with protons in the LS or $\tilde{\nu}_e$ capture by In with the $^{115}$In $\nu_e$ tag. These reactions imply LNV, possibly via a small admixture of wrong-handedness, for example, right-handed currents (RHC) in the weak interaction. In LENS both cross channels are available with sources of $^{12}$N ($\nu_e$: 0-16.3 MeV) and $^6$He ($\tilde{\nu}_e$: 0-3.5 MeV). The sensitivity to LNV depends on the hypothetical base rate calculated as if the $\nu_e$ from $^{12}$N were all $\tilde{\nu}_e$ and the $\tilde{\nu}_e$ from $^6$He were all $\nu_e$. These cross channels can be separated by the tags, i.e. by the ($\tilde{\nu}_e$,p) tag on $\nu_e$'s from $^{12}$N and by the $\nu_e$ tag of In on $\tilde{\nu}_e$'s from $^6$He. Of these two, the former is far more sensitive because: the $\tilde{\nu}_e$ tag has a higher cross section, there are many more protons than In atoms and the energy range is higher so that background is smaller. Thus the $^{12}$N cross channel is preferred. The base rate for a $^{12}$N source ($10^{22}$ $\nu_e$ /y), induced hypothetically by 100% wrong helicity in the ($\tilde{\nu}_e$,p) signal window of 5-14.6 MeV is $0.75 \times 10^6$/y tagged events in LENS.

The experimental limit to admixed $\tilde{\nu}_e$ is set by two main background sources: 1) false events in the signal window from $\tilde{\nu}_e$ contamination either in the production of $^{12}$N (negligible at >5MeV) or from external sources and 2) random coincidences. The external source is the unshieldable atmospheric $\tilde{\nu}_e$ at a flux $\sim 1.5 \times 10^{-3}$/cm$^2$s or $\sim 7.5 \times 10^{-4}$/cm$^2$s due to geomagnetic cut off near the equator. This creates $\sim 3 \times 10^{-7}$ false $\tilde{\nu}_e$ events/y, thus an LNV limit R < ($3 \times 10^{-7}/0.75 \times 10^6$) $\sim 4 \times 10^{-13}$.

Random coincidences are given by N1xN2xτ where N1 is the rate in the 2.2 MeV γ of the ($\tilde{\nu}_e$,p) tag (mixed with signal neutrons and external neutron leakage) and N2, the background rate in the signal window 5-14.6 MeV (mainly from $^{12}$N ν-e scattering) and τ the coincidence time window in which the signal and random events occur at different delays, thus separable. With N1$\sim 10^3$/y, N2$\sim 4.8 \times 10^4$/y, τ$\sim 2 \times 10^{-4}$s ($7 \times 10^{-12}$/ y) the random *time* coincidence rate is $3.4 \times 10^{-4}$/y. This is yet constrained by $10^{-3}$ by the *space* coincidence condition on N1 and N2 (within 30 cm). The total random rate is thus $\sim 3.4 \times 10^{-7}$/y. Even though this rate by itself is unobservably small, the *measured* rates of N1, N2 and the time gate τ can lead to valid background estimates as shown above. The wrong helicity limit from random events is thus R <$4.5 \times 10^{-13}$.

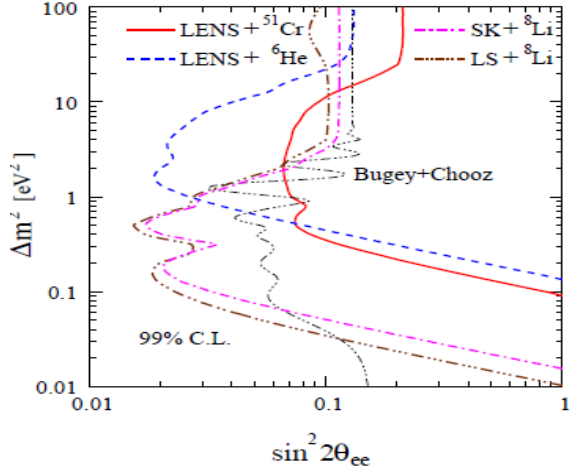

Fig. 2 Sensitivity limits for $P_{ee}$ oscillations (3+1model) of $\nu_e$ (fluence $1.1 \times 10^{24}$) from internal source of $^{51}Cr$ ( In tag) and $\tilde{\nu}_e$ (fluence $2.5 \times 10^{22}$) from $^6$He (($\tilde{\nu}_e$,p) tag ) in LENS (30wt% In). Also shown are results from external (20 m) $\tilde{\nu}_e$ sources ($2.5 \times 10^{22}$) in SK(Gd) and LS-LENA compared to $\tilde{\nu}_e$ limits from Bugey-Chooz. LENS will provide the first $\nu_e$ data.

Overall therefore, R< $8 \times 10^{-13}$ could be expected for $^{12}$N decay. A similar estimate for the conjugate $^6$He channel is R <$5 \times 10^{-11}$. These limits can be put in context of LNV limits set for a variety of processes: R < $10^{-3}$ to <$10^{-9}$ for single β and K decay (Table 10.1[19]) and more recent limits on branching ratios R<$1.7 \times 10^{-12}$ for $\mu^-$-$e^+$ [20] and R< $3 \times 10^{-11}$ for μ-eγ conversions[21]. The role of RHC in 0νββ decay has long been discussed[19]. The partial probability of 0νββ via RHC is[19]: R($\lambda^2$) <$68 \times 10^{-13}$ and a somewhat lower limit for R($\eta^2$) where λ and η are the V+A amplitudes[19]. Thus a limit R<$8 \times 10^{-13}$ for LNV via wrong helicity ν capture in $^{12}$N competes meaningfully with 0νββ. Moreover, interpretation of R(ν capture) is cleaner than R(0νββ) (no nuclear matrix elements or competing processes), promising a sensitive and reliable new probe for LNV.

In summary, we propose a new approach for sensitively probing short baseline oscillations using MeV $\tilde{\nu}_e$ and $\nu_e$ beams produced on line in internal geometries in detectors with specific tags for CC $\tilde{\nu}_e/\nu_e$ capture. The major dividend is observation of the explicit $P_{ee}$ wave within the detector itself for unique proof of oscillations. The approach is versatile enough to access, particularly in LENS, tests of CPT and of LNV comparable to 0νββ-decay. With exclusive tags and signal windows, (a-s) mixing, CPT symmetry and LNV can be tested concurrently with the basic LENS program, viz. ν calibration of the $^{115}$In reaction by a terrestrial $\nu_e$ source and precision measurement of pp and other low energy solar neutrinos from the sun.

We thank J. M. Link, P. Huber (VT), D. J. Thomson (Queens U., Kingston) and M. Hirsch (IFIC, Valencia) for helpful discussions.

* Corresponding author : raghavan@vt.edu
# Present address : Sanjib.Agarwalla@ific.uv.es